\documentclass{article}
\usepackage{spconf,amsmath,graphicx}
\usepackage{cite}
\usepackage{lipsum}
\usepackage{multirow}
\usepackage{amsmath}
\usepackage{amsfonts}
\usepackage{soul}

\title{PROCTER: PROnunciation-aware ConTextual adaptER for personalized speech recognition in neural transducers}
\name{%
\begin{tabular}{@{}c@{}}
Rahul Pandey$^{1, 2}$\sthanks{Work done as an applied scientist intern at Amazon Alexa.} \qquad
Roger Ren$^{1}$\sthanks{Contributed equally.} \qquad
Qi Luo$^{1}$\footnotemark[2] \qquad
Jing Liu$^{1}$\footnotemark[2] \qquad
Ariya Rastrow$^{1}$ \quad \\
Ankur Gandhe$^{1}$ \quad
Denis Filimonov$^{1}$ \quad
Grant Strimel$^{1}$ \quad
Andreas Stolcke$^{1}$ \quad
Ivan Bulyko$^{1}$ \quad
\end{tabular}}
\address{$^1$Amazon Alexa AI, USA \\
$^2$George Mason University, USA }
\begin{document}
\ninept
\maketitle
\begin{abstract}
End-to-End (E2E) automatic speech recognition (ASR) systems used in voice assistants often have difficulties recognizing infrequent words personalized to the user, such as names and places. Rare words often have non-trivial pronunciations, and in such cases, human knowledge in the form of a pronunciation lexicon can be useful.
We propose a \textbf{PRO}nun\textbf{C}iation-aware con\textbf{T}extual adapt\textbf{ER} (PROCTER) that dynamically injects lexicon knowledge into an RNN-T model by adding a phonemic embedding along with a textual embedding. 
The experimental results show that the proposed PROCTER architecture outperforms the baseline RNN-T model by improving the word error rate (WER) by $44\%$ and $57\%$ when measured on personalized entities and personalized rare entities, respectively, while increasing the model size (number of trainable parameters) by only $1\%$. Furthermore, when evaluated in a zero-shot setting to recognize personalized device names, we observe $7\%$ WER improvement with PROCTER, as compared to only $1\%$ WER improvement with text-only contextual attention. 
\end{abstract}


\section{Introduction}
\label{sec:intro}

End-to-end (E2E) ASR systems \cite{chan_listen_2016,dong2018speech,graves2012sequence, Yeh2019TransformerTransducerES} 
have achieved significant WER improvements over traditional hybrid systems. However, they often have difficulty in correctly recognizing words that appear infrequently in the training data.  Entity lists personalized to specific users (e.g., personalized entity lists or user-defined personalized device names) can help improve ASR accuracy for virtual voice assistants. 
\begin{figure}[t]
\centering
  \centerline{\includegraphics[width=6cm]{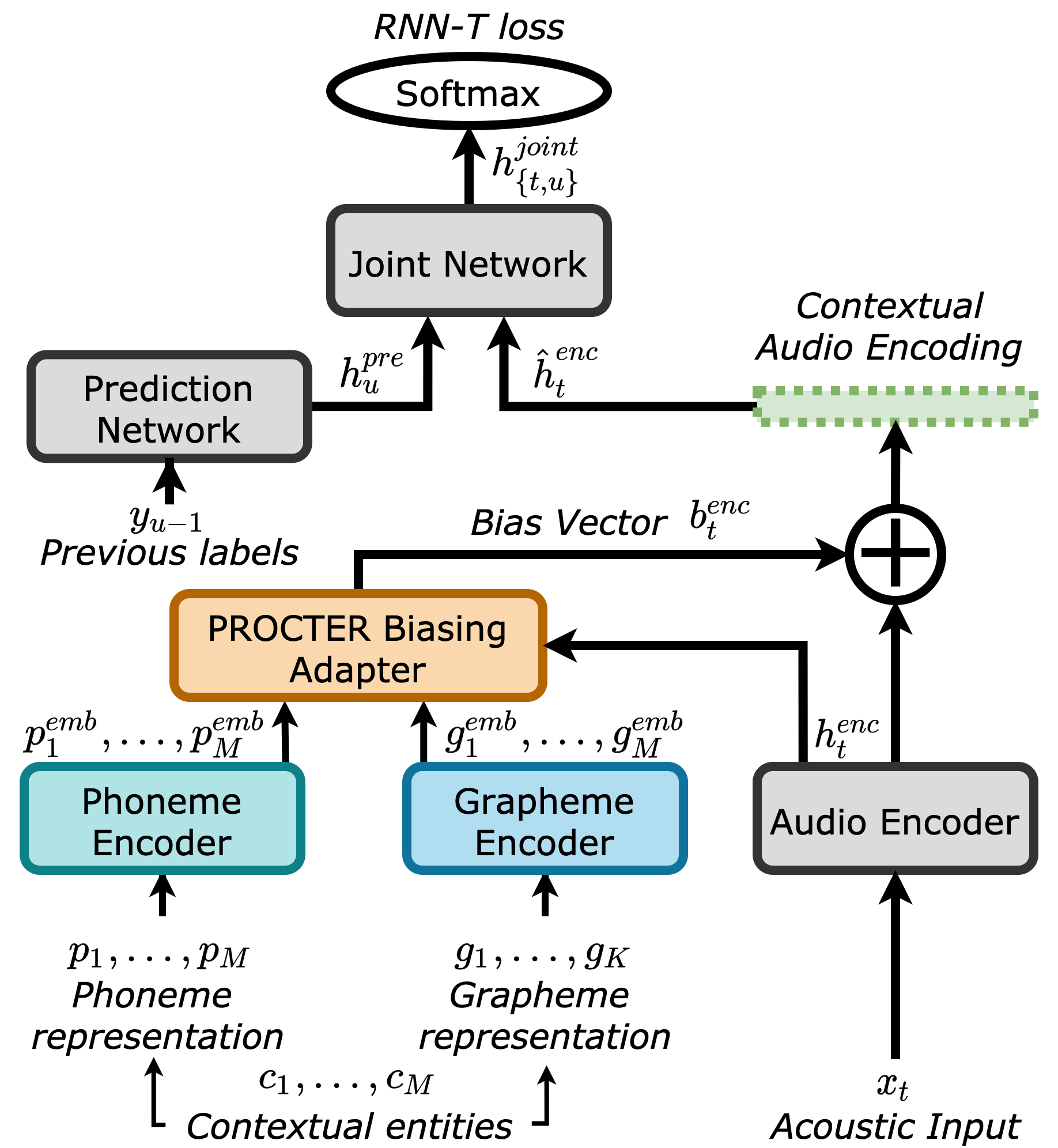}}
\caption{Architecture of \textbf{PROCTER}.}
\label{fig:archiecture}
\vspace{-6.5mm}
\end{figure}

\begin{figure*}[htb]
\centering
  \centerline{\includegraphics[width=0.9\linewidth]{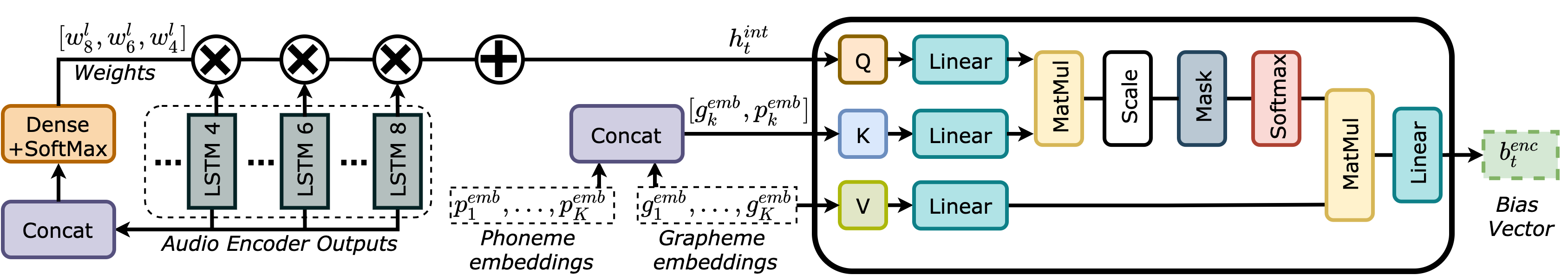}}
\caption{Proposed PROCTER biasing adapter}
\label{fig:intermediate}
\vspace{-6.5mm}
\end{figure*}
There has been much recent prior work to improve recognition of rare words for E2E ASR systems by incorporating additional contextual information \cite{sathyendra2022contextual,chang2021context,bruguier2019phoebe,pundak2018deep,jain2020contextual,sriram_cold_2018,renduchintala_multi-modal_2018,kannan_analysis_2018,karita_semi-supervised_2018,bruguier_learning_2016,gourav_personalization_2021,le_contextualized_2021,le_deep_2021}. 
The contextual information is ingested either post-training (e.g., by shallow fusion \cite{gourav_personalization_2021, zhao_shallow-fusion_2019} and on-the-fly rescoring \cite{he2019streaming, yang2021multi, xu2022rescorebert}) or during training (e.g., with contextual adapters \cite{chang2021context}). However, the efficacy of the post-training method is limited due to lack of joint training with the core recognition components, making these approaches sensitive to heuristics used for tuning their rescoring weights. In contextual adapter approaches, however, context is directly and jointly used as part of the recognition (e.g., RNN-T) model to improve the E2E ASR loss.

Most prior work leverages only the textual representation when incorporating contextual information. However, given that contexts like personalized entities are infrequent during training and often have multiple pronunciations, the E2E ASR model can get confused by similar textual contexts. 
For example, \textit{``Jana''} has a pronunciation \textit{``Yana''} that is not evident from the graphemes (in English).  Thus, adding pronunciation information would make it easier for the model to align the entity \textit{``Jana''} to the input audio, and hence infer the correct output. There is little prior work that utilizes phonemic representations of context, via a simple additive attention mechanism \cite{bruguier2019phoebe, chen_joint_2019} that is specific to the LAS\cite{chan_listen_2016} E2E architecture.

In this paper, we propose a jointly-trained contextualization adapter for the RNN-T model by incorporating both textual (graphemic) and phonemic representations of context to create a pronunciation-aware contextual adapter (\textbf{PROCTER}). First, given the contextual entities, we encode its grapheme and phoneme representation to get the grapheme and phoneme embeddings, respectively. Then, they are used to bias the intermediate outputs from the audio encoder of the RNN-T model to fuse the contextual information using the PROCTER biasing adapter before passing it to the joint network of RNN-T. In contrast to previous approaches that select one pronunciation per contextual entity \cite{bruguier2019phoebe, chen_joint_2019}, we preserve all pronunciations by considering them as separate entities and duplicate the corresponding textual representation in multiple grapheme-phoneme pairs. Additionally, unlike the attention mechanism used in prior work  \cite{bruguier2019phoebe}, our adapter biases the audio encoder outputs with contextual information using a more optimized scaled dot-product attention \cite{vaswani2017attention}. Prior work has shown that the output corresponding to the last layer of the audio encoder in E2E ASR models has less phonemic information since ASR is optimized to output the correct graphemes irrespective of phonemic variations \cite{pasad2021layer, pmlr-v139-yang21j}. Thus, instead of letting only the last layer attend to contextual entities \cite{bruguier2019phoebe, chang2021context}, we also bias earlier layers with our phonemic embeddings. Providing phonemic context allows the biasing network to pay attention to idiosyncratic pronunciation variants. Moreover, allowing multiple grapheme-phoneme pairs lets the model attend to the best-matching pronunciation of a context item. We train the model in adapter style, freezing the rest of the RNN-T model with weights from pretraining. Thus, it requires a small fraction of the data and time needed for full training. To demonstrate the performance of our proposed model, we compared PROCTER with a state-of-art model (i.e., a text-only) contextual adapter model \cite{sathyendra2022contextual, chang2021context}) on three in-house far-field datasets.

\section{Prior Work}
\subsection{Neural Transducer}
\label{sec:rnn-t}
The recurrent neural transducer (RNN-T) is widely used as an E2E ASR streaming model. The grey blocks in Figure~\ref{fig:archiecture} represent the streaming RNN-T ASR system, which has three main components: an audio encoder, a prediction network, and a joint network. 

The audio encoder typically consists of stacked LSTM layers \cite{graves2012sequence}. It generates the high-level audio encoding representations $h^{enc}_t$ given the input audio frames $x_{0,t} = (x_0, ..., x_t)$. 
The prediction network also uses stacked LSTM layers to encode the last nonblank word pieces $y_{0,u-1} = (y_0, ..., y_{u-1})$, and generate $h^{pre}_u$. 
The joint network takes the output from both audio encoder $h^{enc}_t$ and prediction network $h^{pre}_u$. It fuses the two outputs before passing them to a series of dense layers (denoted by $\phi$). Finally, a softmax operation is applied to obtain the probability distribution over word pieces. Equation~\ref{equ:rnnt} shows that RNN-T outputs probability at time $t$ given a sequence $u$. 
\begin{equation}
\label{equ:rnnt}
\vspace{-0.5mm}
\begin{gathered}
h^{joint}_{t,u} = \phi(\text{JoinOp}(h^{enc}_t, h^{pre}_u))
\\
P(y_u\:|\:t, u) = \text{softmax}(h^{joint}_{t,u})
\end{gathered}
\end{equation}
\subsection{Contextual adapters}
\label{sec:con_adap}
Contextual adapters have been shown effective in adapting pretrained RNN-T model for the recognition of contextual entities, achieving up to $31\%$ relative word error rate reduction (WERR) compared to the vanilla RNN-T model \cite{chang2021context,sathyendra2022contextual}. The biasing network has two main components---a catalog encoder and a biasing adapter.

The catalog encoder embeds a catalog of contextual entities $C = {c_1, c_2,..., c_K}$. It takes the graphemic (textual) representation of catalog entities as input, passing them to a subword tokenizer \cite{sennrich2016neural}, followed by a stack of BiLSTM layers. The last state of the BiLSTM is used as the embedding of a contextual entity, which is an encoded representation $c_e$. Given a catalog with $K$ entities, and the generated entity embedding size of $D$, the catalog encoder outputs $C^e \in \mathbb{R}^{K\times D}$ as follows: 
\begin{equation}
\vspace{-0.5mm}
    c^{e}_i = \text{BiLSTM}(Embedding(c_i))
    \label{eqn:cat_enc}
\end{equation}

The biasing adapter transforms an intermediate representation computed by the RNN-T, such as the audio encoder output ($h^{enc}_t$), using the graphemic entity embeddings from the personalized catalog. It uses a cross-attention-based biasing adapter to attend over all entity embeddings $C^{e}$ (key and value) based on $h^{enc}_t$ as the input query. The attention scores $\alpha_i$ for each catalog entity are calculated using the scaled dot-product attention mechanism \cite{vaswani2017attention}.
The biased intermediate representation (bias vector) is calculated as $b^\textit{enc}_t = \sum^{K}_{i} \alpha_i W^v c^e_i$. Finally, the intermediate representations are updated with the bias vector using element-wise addition. Thus, the contextual audio encoding ($\hat{h}^{enc}_t$) that contains the biased representation is calculated as $\hat{h}^\textit{enc}_t = h^\textit{enc}_t \oplus b^\textit{enc}_t$.

One major shortcoming in designing the contextual adapters is its heavy reliance on the textual representation of the catalog entities. While the contextual adapter can learn to effectively represent rare catalog entities, it struggles to encode entities with uncommon pronunciation. Prior work \cite{bruguier2019phoebe,Papadourakis2021} showed that the phonemic representation of such rare entities can be compared more effectively to grapheme-only representations in the case of a contextual LAS model \cite{pundak2018deep}. These findings motivate combining phonemic and graphemic representations within contextual adapters to improve accuracy on rare entities. 
\section{PROCTER}
\label{sec:PROCTER}
The proposed PROCTER model takes as input the phonemic representation of catalog entities in addition to its textual grapheme representation. Figure \ref{fig:archiecture} shows the overview of the architecture. It has three components: phoneme encoder, grapheme encoder, and PROCTER biasing adapter. The objective of PROCTER is to learn the phonemic variations of catalog entity representations such that it improves the biasing of intermediate audio encodings for the contextual catalog entities with nonstandard pronunciations. 
\subsection{Phoneme Encoder}
The phoneme encoder embeds the pronunciation representation of entities in the catalog. The pronunciation information is retrieved from the dictionary. Same-text entities can have multiple pronunciations, which are all preserved as separate entities for biasing the intermediate output representation. Let $P = [p_1, p_2, ... , p_M]$ denote all pronunciations (phoneme sequences) of catalog entities $C = [c_1, c_2, ...., c_K]$ where $M \geq K$. The phoneme encoder embeds the phoneme sequences using an embedding lookup, followed by a stack of BiLSTM layers. With $M$ pronunciations and an embedding size of $D_p$, the phoneme encoder outputs $P^\textit{emb} \in \mathbb{R}^{M\times D_p} $ where $p^\textit{emb}_m \subset P^\textit{emb}$ is the phoneme embedding of pronunciation $p_m$, calculated as
$$
p^{emb}_j = \text{BiLSTM}(Embedding(p_j)) \quad .
$$
We also add a $no\_bias$ token to the pronunciation catalog for disabling adapter biasing, similar to prior work \cite{bruguier2019phoebe, sathyendra2022contextual}. 
\subsection{Grapheme Encoder}
Our grapheme encoder is similar to the catalog encoder described in Section~\ref{sec:con_adap}. It encodes the grapheme representation $G = [g_1, g_2, ..., g_K]$ of catalog entities $C = [c_1, c_2, ...., c_K]$ using Equation~\ref{eqn:cat_enc}. 
The grapheme encoder also duplicates the output grapheme embeddings for the entities with more than one pronunciation to match up with the phoneme embeddings for the various pronunciations of a given entity. Hence, given a total of $M$ pronunciations and grapheme embedding size $D_g$, the grapheme encoder outputs $G^\textit{emb}\in \mathbb{R}^{M \times D_g}$ where $g^\textit{emb}_m \subset G^\textit{emb}$ is the grapheme embedding of the textual representation in the catalog with $p_m$ as one of its pronunciations.  
\subsection{PROCTER Biasing Adapter}
The proposed PROCTER biasing adapter adapts an intermediate representation from the audio encoder with both textual and pronunciation representations of contextual catalog entities, as shown in Figure~\ref{fig:intermediate}. It is based on cross-attention using the scaled dot-product \cite{vaswani2017attention}. The cross-attention module of the proposed adapter takes three inputs: query, key, and value. 

\textit{Query}: The query comes from the intermediate layer output of the pretrained audio encoder. Instead of using only the final layer output, we also include intermediate layer outputs earlier in the pretrained RNN-T audio encoder. More specifically, we utilize the outputs corresponding to the last, third-last, and fifth-last LSTM layers. We compute weighted addition of these encoder outputs as a gating mechanism. Given the intermediate layer outputs $\{ h^{\textit{enc}^{l}}_t, h^{\textit{enc}^{l-2}}_t, h^{\textit{enc}^{l-4}}_t \}$, we use a dense layer to project the concatenated intermediate layer outputs to a 3 dimensional vector for weights $[w^{l}_t, w^{l-2}_t, w^{l-4}_t]$. And these weights are used to calculate the final weighted sum of encoder output  $h^\textit{qry}_t$, which is used as the query to the scaled dot-product attention as shown in Figure~\ref{fig:intermediate}.
\vspace{-1.5mm}
\begin{equation}
\label{eqn:query}
\begin{gathered}
h^{enc^{cat}}_t = \text{Concat}(\{h^{\textit{enc}^{l}}_t, h^{\textit{enc}^{l-2}}_t, h^{\textit{enc}^{l-4}}_t\}) 
\\\vspace{-1.5mm} 
[w^{l}_t, w^{l-2}_t, w^{l-4}_t] = \text{Softmax}(\text{Dense}(h^{\textit{enc}^\textit{cat}}_t))
\\
\\h^\textit{qry}_t = \sum_{i \in \{l, l - 2, l- 4\}} w^{i}_t h^{\textit{enc}^{i}}_t
\end{gathered}
\end{equation}

\textit{Key}: The key is used to compute the attention score based on the provided query. We use both textual and pronunciation representations coming from the grapheme and phoneme encoders, respectively, as the key. Since we have a one-to-one mapping of corresponding phoneme and grapheme embeddings, we concatenate them to obtain the final embedding for the key. For each catalog grapheme-phoneme embedding pair $g^\textit{emb}_m$ and $p^\textit{emb}_m$ where $m \in {1,\ldots, M}$, we compute the input to key $c^\textit{key}_m$ as
\begin{equation}
\label{equ:concat_key}
    c^\textit{key}_m = \text{Concat}(g^\textit{emb}_m, p^\textit{emb}_m)
\end{equation}
Thus, the key for all grapheme-phoneme embedding pairs becomes $C^\textit{key} = [c^\textit{key}_1, c^\textit{key}_2, ..., c^\textit{key}_M]$.

\textit{Value}: The value is used to bias the query with the computed attention weights. Since the biased vector passed to the joint network of RNN-T is expected to contain only textual information, we use the textual representation $g^\textit{emb}_m$ as the value. 

We compute an attention score $\alpha_i$ based on the weighted sum of intermediate encoder outputs (Eqn.~\ref{eqn:query}) as query and concatenated grapheme-phoneme representations of all catalog entities as the key: 
\begin{equation}
     \alpha_i = Softmax_i\left ( \frac{W^qh^{qry}_t\cdot \left (W^kC^{key} \right )^T}{\sqrt{d}} \right ) 
    \label{eqn:bias_adap}
\end{equation}
The final bias vector $b^\textit{enc}_t$ is calculated as $b^\textit{enc}_t = \sum^{K}_{i} \alpha_i W^v g^\textit{emb}_i$. Finally, similar to \cite{sathyendra2022contextual}, the intermediate representations are fused with the contextual bias vector using element-wise addition. Thus, the contextual audio encoding ($\hat{h}^\textit{enc}_t$) that contains the biasing contextual representation is calculated as $\hat{h}^\textit{enc}_t = h^\textit{enc}_t \oplus b^\textit{enc}_t$.

The intuition behind the proposed PROCTER strategy is to learn better attention weights by including the phonemic representation of the catalog entities. Moreover, to better index into the phonemic representation, we also include the intermediate layer outputs of the pretrained audio encoder to capture acoustic-phonetic variations in the  audio inputs.
\section{Experiments}
\begin{table*}[ht]
\centering
\caption{\textbf{Results.} Relative change in WER (WERR), and NE-WER (NE-WERR) over vanilla RNN-T models for various models. The rare personalized entities appear only once, and personalized device name test set shows the zero-shot capability of the model.}
\tabcolsep=0.03cm
\begin{tabular}{|l|cc|c|c|ccc|cc|}
\hline
\multirow{3}{*}{Model}                                                                    & \multicolumn{2}{c|}{\multirow{2}{*}{Phoneme}}    & \multirow{3}{*}{Int-Layers} & \multirow{2}{*}{General} & \multicolumn{3}{c|}{Personalized Entity}                                                                                                                                                         & \multicolumn{2}{c|}{\begin{tabular}[c]{@{}c@{}}Personalized Device Name\end{tabular}}            \\ \cline{6-10} 
                                                                                                  &  &        &        &          & \multicolumn{1}{c|}{Overall}          & \multicolumn{1}{c|}{\begin{tabular}[c]{@{}c@{}}All Entities \end{tabular}} & \begin{tabular}[c]{@{}c@{}}Rare Entities \end{tabular} & \multicolumn{1}{c|}{Overall}         & \begin{tabular}[c]{@{}c@{}}Personalized Device Names \end{tabular} \\ \cline{2-3} \cline{5-10} 
                                                                                               & \multicolumn{1}{c|}{key} & value  &   & WERR                     & \multicolumn{1}{c|}{WERR}             & \multicolumn{2}{c|}{NE-WERR}                                                                                                                      & \multicolumn{1}{c|}{WERR}            & NE-WERR                                                     \\ \hline
\textit{Text-only adapter}                                                                & \multicolumn{1}{c|}{N} & N       & N           & 0.0\%                    & \multicolumn{1}{c|}{32.5\%}          & \multicolumn{1}{c|}{37.1\%}                                                      & 50.0\%                                                       & \multicolumn{1}{c|}{1.1\%}          & 0.9\%                                   \\ \hline \hline
PROCTER                                                                                  & \multicolumn{1}{c|}{Y} & N   & Y         & 0.2\%                   & \multicolumn{1}{c|}{\textbf{38.2\%}} & \multicolumn{1}{c|}{\textbf{43.9\%}}                                             & \textbf{57.2\%}                                              & \multicolumn{1}{c|}{\textbf{2.6\%}} & \textbf{6.9\%}                    
\\ \hline \hline
\textit{PROCTER+Ph-in-value}                                                                 & \multicolumn{1}{c|}{Y} & Y      & Y     & 0.2\%                   & \multicolumn{1}{c|}{\textbf{38.2\%}} & \multicolumn{1}{c|}{43.2\%}                                                      & 55.1\%                                                       & \multicolumn{1}{c|}{1.3\%}          & 5.3\%                                                      \\ \hline
\begin{tabular}[c]{@{}c@{}}\textit{PROCTER+No-Int-Layers}\end{tabular}        & \multicolumn{1}{c|}{Y} & N     & N       & \textbf{1.0\%}          & \multicolumn{1}{c|}{\textbf{38.2\%}} & \multicolumn{1}{c|}{43.6\%}                                                      & 57.1\%                                                       & \multicolumn{1}{c|}{0.2\%}          & 3.9\%                                                      \\ \hline
\begin{tabular}[c]{@{}c@{}}\textit{PROCTER+Ph-in-value+No-Int-Layers}\end{tabular}& \multicolumn{1}{c|}{Y} & Y & N  & 0.2\%                   & \multicolumn{1}{c|}{37.6\%}          & \multicolumn{1}{c|}{43.0\%}                                                      & 54.6\%                                                       & \multicolumn{1}{c|}{0.0\%}          & 4.8\%                                                                                      \\ \hline
\end{tabular}

\label{tab:res}
\vspace{-3.5mm}
\end{table*}
\subsection{Dataset and Evaluation}
We use in-house de-identified far-field datasets coming from interactions with a virtual voice assistant. The training data consists of text-audio pairs of utterances randomly sampled from more than 20 domains such as Communications, Weather, SmartHome, and Music. The baseline RNN-T model was trained using 114k hours of data. However, following the work in \cite{sathyendra2022contextual,chang2021context}, the attention components in PROCTER are trained using a fine-tuning step where the core components of RNN-T (encoder, prediction) are kept frozen. For training the adapter, we use much less, $\sim 290$ hours of data. Personalized entities from real users are used as context information. We test the model on three test sets: 1) 16k utterances of far-field English data, 2) 29k utterances of far-field English data, which includes mentions of the personalized entities, 3) 3558 utterances of data containing mentions of personalized device names. We report the WERR on all three test sets, as well as the relative named-entity WER reduction (NE-WERR) of personalized entities and personalized device names, respectively, on the second and third test sets. For the second test set, we further compute the NE-WERR of rare personalized entities (those appearing only once in the test set), to specifically measure performance on infrequent words.  

\subsection{Experimental Setup}
\noindent \textbf{Pretrained RNN-T model.} We use the RNN-T model described in Section \ref{sec:rnn-t}. The input is a 64-dim LFBE feature for every 10\,ms of audio, with a window size of 25\,ms, with three frames stacked together resulting in 192 features per frame. Each ground truth text token is passed through a $4000$ word-piece tokenizer \cite{sennrich2016neural}. The RNN-T audio encoder consists of 8 LSTM layers with 1280 units per layer.
The prediction network consists of 2 LSTM layers with 1280 units per layer. The joint network consists of a dense layer of 512 units, followed by a softmax activation over RNN-T output tokens. The decoding is performed using the standard beam search with a beam size of 8. The output vocabulary consists of 4000 word pieces.

\noindent \textbf{PROCTER configuration.} The phoneme sequences of contextual entities are generated from a lexicon. A text can have multiple pronunciations, and we found it important to keep all phoneme sequences. 
The grapheme and phoneme encoders consist of BiLSTMs with 64 and 128 units, respectively. 
The PROCTER biasing adapter projects the query, key, and value to 128 dimensions with attention weights. The maximum number of contextual entities is set to 300. For the PROCTER-based experiments, we further divide the full personalized entities or personalized device names into individual words. The words are duplicated based on their number of pronunciations in the lexicon file for a one-to-one correspondence during concatenation. We keep a maximum size of 600 for these phoneme-grapheme pairs. For training the PROCTER, we use the Adam optimizer with a learning rate of $5 \times 10^{-4}$ configured to converge with early stopping. The number of trainable adapter parameters is 1.5M, or about 1\% of the RNN-T. 

\noindent \textbf{Experiment setup.} 
The primary baseline model is a vanilla RNN-T model without personalization described in Section~\ref{sec:rnn-t}.   We compare the proposed PROCTER adapter to four additional model configurations:
	1) \textbf{\textit{Text-only adapter}}: Contextual adapter proposed in \cite{sathyendra2022contextual} using only text representations of context entities;
	2) \textbf{\textit{PROCTER + Ph-in-value}}: PROCTER with the value embeddings created the same way as the keys (Equation~\ref{equ:concat_key}). This experiment assesses the effect of using phonemic information in the final contextual embedding that is passed into the joint network;
	3) \textbf{\textit{PROCTER + No-Int-Layers}}: PROCTER with the queries defined using the final layer output of the pretrained audio encoder (as opposed to using the weighted sum of intermediate outputs). This ablation study examines the importance of information from intermediate audio encoder layers when attending to the contextual embeddings;
	4) \textbf{\textit{PROCTER + Ph-in-value + No-Int-Layers}}: Combination of variants (2) and (3).

\section{Results}

Table 1 shows the results relative to the vanilla RNN-T model on all three test sets. Given the general test set with no personalized context, we observe that the previous \textit{text-only adapter}, proposed PROCTER, and all ablation experiments have no degradation over the baseline RNN-T model. It shows that the contextual adapters are immune to the performance of general ASR. For the personalized entity test set, in terms of overall WERR, we observe that all the experiments significantly improved over the vanilla RNN-T, which shows the effectiveness of contextual adapters. However, we see a gain when we introduce the phonemic context representation compared to the \textit{text-only adapter} (38.2\% vs. 32.5\%). Furthermore, the performance gain increases compared to the \textit{text-only adapter} as the sparsity of context increases, as seen in the NE-WERR of all personalized entities and rare personalized entities (43.9\% \& 57.2\% vs.\ 37.1\% \& 50\%). Although all the ablation experiments perform similarly to the PROCTER model in overall WERR and NE-WERR for all personalized entities,
we see a performance gain for PROCTER on rare personalized entities over the first and third ablation experiments (57.2\% vs.\ 55.1\% and 54.6\%). This shows the importance of not passing the phoneme information to the joint network of the RNN-T model. 

Additionally, we test the zero-shot capability of PROCTER. We test the personalized device name test set with personalized device names as context, with all the experiments trained with personalized entities for contextualization. Since the context domain differed, we did not see much improvement over the vanilla RNN-T model. However, we observe the highest performance gain by PROCTER in terms of overall WERR and personalized device name NE-WERR. In particular, we achieved 6.9\% NE-WERR of the personalized device name context compared to 1.0\% with \textit{text-only adapter}. Moreover, we also observe better performance of PROCTER with the second ablation experiment. Therefore, when the sparsity of context is at the maximum, it is essential to include the intermediate layer audio encoder outputs for better contextualization. 

Finally, to demonstrate the learning of phonemic context representation, Figure~3 shows an example (the personalized entity was altered to protect privacy) where the PROCTER model identifies the correct entity while the \textit{text-only adapter} fails. \textit{Text-only adapter} attends to another textually similar personalized entity, `baden', for the audio frames corresponding to `biden', leading to an incorrect hypothesis, ``call joe baden''. However, the PROCTER model, with access to the pronunciations of personalized entities `biden' and `baden', identifies the context `biden' to arrive at the correct ASR hypothesis. 

\begin{figure}[h]
\label{fig:heatmap}
\begin{minipage}[b]{\linewidth}
\vspace{-4mm}
  \centering
  \centerline{\includegraphics[width=6.0cm]{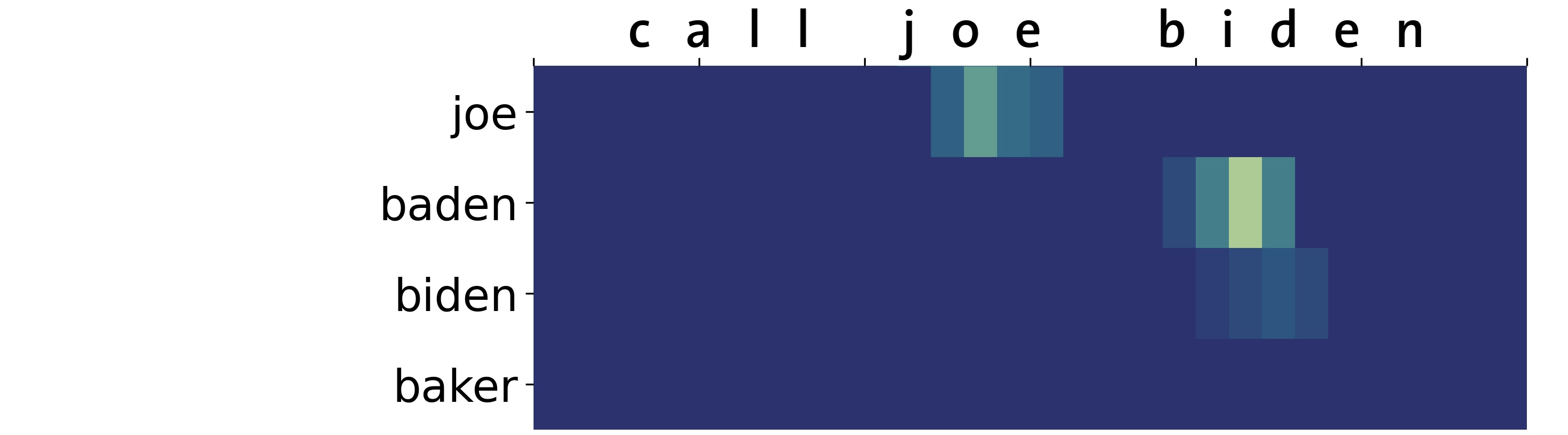}}
  \centerline{(a) \textit{text-only adapter}}\medskip
   \vspace{-2mm}
\end{minipage}
\begin{minipage}[b]{\linewidth}
  \centering
  \centerline{\includegraphics[width=6.0cm]{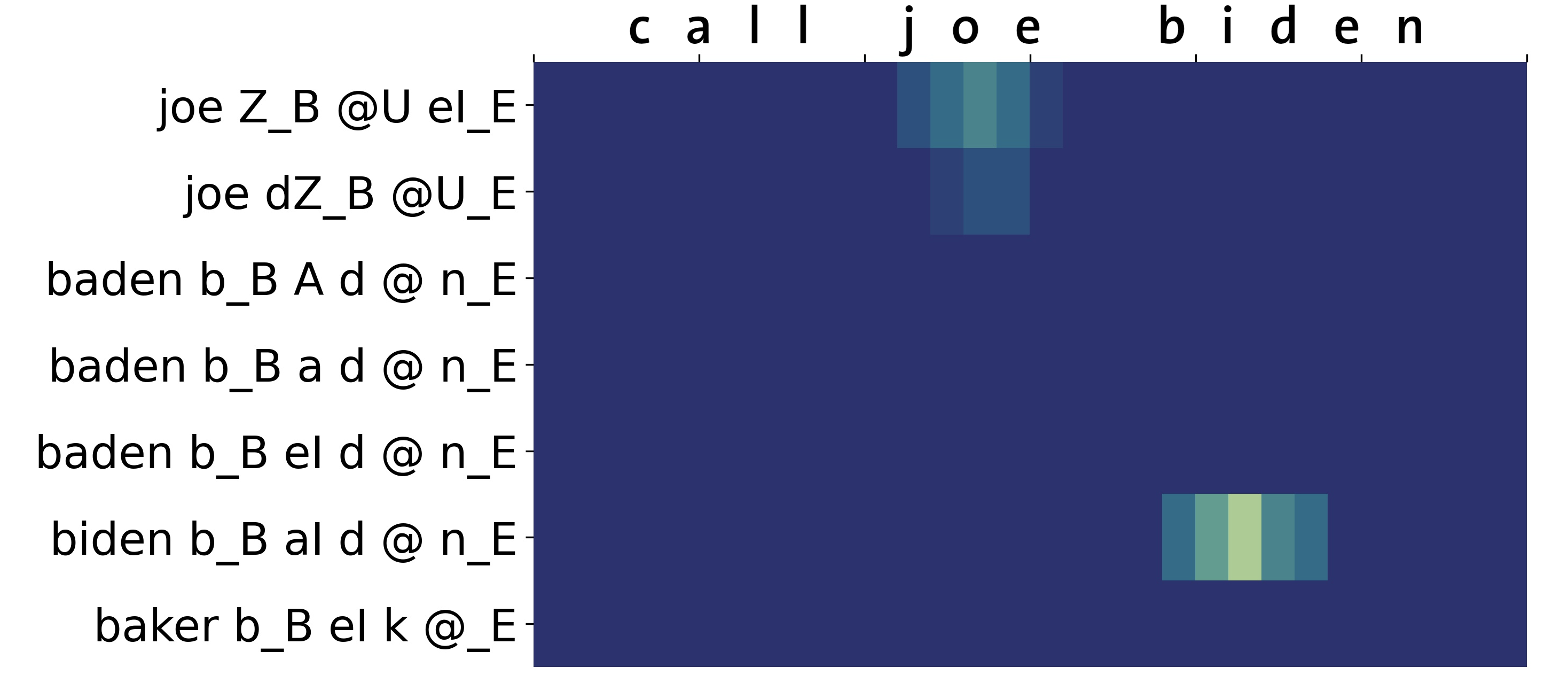}}
  
  \centerline{(b) PROCTER}\medskip
 \vspace{-2mm}
\end{minipage}

\vspace{-3.2mm}
\caption{Attention visualization}
\label{fig:res}
\vspace{-5mm}
\end{figure}
\vspace{-3.5mm}
\section{Conclusion}
\vspace{-2mm}
\label{sec:conclusion}
We proposed PROCTER, a pronunciation-aware contextual adapter for a vanilla RNN-T model for improving the recognition of rare words. It fuses the pronunciation representations with the textual representations of context words via scaled dot-product attention. PROCTER is trained in adapter style with only 1\% of the trainable parameters of a vanilla RNN-T model. As a result, we observe over 57\% improvement compared to the baseline RNN-T model on rare personalized entities and up to 7\% improvement in zero-shot testing on a personalized device name test set. Although PROCTER relies on a dictionary for pronunciations, we found less than 0.5\% of contexts lacking pronunciation information. In the future, we plan to use a neural grapheme-to-phoneme model for those rare cases. We also plan to try applying PROCTER-style biasing on decoder outputs in addition to audio encoder outputs.

\footnotesize 
\bibliographystyle{IEEEbib}
\bibliography{refs}

\end{document}